\documentclass{article}%
\usepackage{graphics}
\usepackage{eurosym}
\usepackage[top=3cm, bottom=3cm, left=3cm, right=3cm]{geometry}
\usepackage{amsmath}
\usepackage{amsfonts}
\usepackage{amssymb}
\usepackage{graphicx}%
\usepackage{epsfig}
\begin{document}

\title{Mass spectra and transition magnetic moments of low lying charmed baryons in a quark model}
\author{Zahra Ghalenovi  \footnote{$z_{-}ghalenovi@kub.ac.ir$} and Masoumeh Moazzen Sorkhi\\Department of Physics, Kosar University of Bojnord, Iran
}
\maketitle
\begin{abstract}
\noindent Excited state mass spectra of the low lying single charmed baryons with non strangeness have been calculated in a hypercentral approach. The six-dimensional hyperradial Schr\"{o}dinger  equation is solved by applying a simple variational method. We extend our scheme to predict the magnetic moments and the  $ \frac{3}{2}^+ \rightarrow  \frac{1}{2}^+ $ transition magnetic moments of $ \Sigma_{c} $  and $ \Lambda_{c} $  state baryons. A comparison of our results with the experimental data and predictions obtained in recent models is also presented.\\


\end{abstract}

\section{ Introduction}

The $ \Sigma_{c}(2455) $  and $ \Lambda_{c}(2286) $ are the two lowest baryon states in the heavy flavor sector. Their descriptions and properties thus play a very important role in the understanding of strong interaction.  Many theoretical models such as non-relativistic model \cite{Shah:42, Ghalenovi:2013}, relativistic quark model \cite{Ebert5}, Lattice QCD \cite{Mathur}, QCD sum rule \cite{Mao, Liu3, Aliev6, Aliev7}, Quark-diquark model \cite{Ebert6, Thakkar6, Thakkar7} and Faddeev approach \cite{Gerasyuta} also have been presented to study the properties of heavy flavor baryons, however there are limited numbers of theoretical studies on the excited states and transition magnetic moments of heavy baryons. 
The purpose of this paper is to study the properties of non-strange single charm baryons. We calculate the ground and excited states mass spectra and magnetic moments of $ \Sigma_{c} $  and $ \Lambda_{c} $  baryons. We calculate the baryon spectrum in a two-step procedure: first, we introduce the perturbating hyperfine interaction and then obtain the baryon masses. We also compute the spin-orbit potential for the excited baryons. Second, we use the hypercentral model in order to solve the three-body Schr\"{o}dinger equation by performing an variational method and obtain the wavefunction and eigenvalues of the baryon systems. \\
The measurement and calculation of electromagnetic transitions $ {B\rightarrow B^\prime \gamma}$ between $J=3/2^{+}$ and $J=1/2^{+}$ baryons is an important issue to understand the internal structure of baryons and observe new hadronic states experimentally. Here, we extend our scheme to calculate the $ \frac{3}{2}^+ \rightarrow  \frac{1}{2}^+ $ transition magnetic moments of single charm baryons in a simple way.
The paper is organized as follows. In Sec.\ 2 we introduce the hypercentral model. We present  our potential model to solve the Schr\"{o}dinger equation in Sec. \ 3. Our predictions for the baryon masses, magnetic moments and $ \frac{3}{2}^+ \rightarrow  \frac{1}{2}^+ $ transition magnetic moments are presented in Sec. 4  and Sec.\ 5 includes conclusions.\\

\section{The hypercentral constituent quark model}
\subsection{Hyperspherical coordinates}
We consider the baryon system as a bound state of three constituent quarks. After removal of the center of mass coordinates, $ R $, the configurations of three quarks can be described by the Jacoobi coordinates, $ \rho $ and $ \lambda $,
\begin{equation}  \label{rhoo}%
\vec{\rho} = \frac{1}{\sqrt{2}}(\vec{r_1} - \vec{r_2}), \qquad
\vec{\lambda} = \frac{1}{\sqrt{6}}(\vec{r_1} + \vec{r_2} - 2
\vec{r_3})  
\end{equation}
such that
\begin{equation}
m_{\rho} = \frac{2 m_1 m_2}{m_1 + m_2}, \qquad m_{\lambda} =
\frac{3 m_3 (m_1 + m_2)}{2 (m_1 + m_2 + m_3)}  
\end{equation}
Here $m_1$, $m_2$ and $m_3$ are the constituent quark masses.
Instead of $\rho$ and $\lambda$ , one can introduce the
hyperspherical coordinates, which are given by the angles
$\Omega_{\rho}=(\theta_{\rho},\phi_{\rho})$ and $\Omega_{\lambda}=(\theta_{\lambda},\phi_{\lambda})$ together with the
hyperradius $x$, and the hyperangle, $\zeta$ defined in terms of the absolute values of   $\rho$ and $\lambda$  
by 
\begin{equation}
x = \sqrt{\rho^2 + \lambda^2}, \qquad \xi = \arctan
(\sqrt{\frac{\rho}{\lambda}})
\end{equation}

Therefore the Hamiltonian will be
\begin{equation}
H=\frac{p^{2}_{\rho}}{2m}+\frac{p^{2}_{\lambda}}{2m}+V(x).
\end{equation}

In hyperspherical coordinates the Laplace operator for three-body system is written as follows:
\begin{equation}    \label{laplace}
\nabla^{2}=(\nabla^{2}_{\rho}+\nabla^{2}_{\lambda})=-(\frac{d^{2}}{dx^{2}}+\frac{5}{x}\frac{d}{dx}-\frac{L^{2}(\Omega_{\rho},\Omega_{\lambda},\xi)}{x^{2}}),
\end{equation}

Therefore the kinetic energy operator of a three-body problem can be written as ($\hbar=c=1$):
\begin{equation}
-\frac{1}{2m}(\nabla^{2}_{\rho}+\nabla^{2}_{\lambda})=-\frac{1}{2m}(\frac{d^{2}}{dx^{2}}+\frac{5}{x}\frac{d}{dx}-\frac{L^{2}(\Omega_{\rho},\Omega_{\lambda},\xi)}{x^{2}}).
\end{equation}

The eigenfuctions of the  grand angular operator $L^{2}(\Omega)$ are the so called hyperspherical harmonics
\begin{equation}
L^{2}(\Omega_{\rho},\Omega_{\lambda},\xi)Y_{[\gamma],l_{\rho},l_{\lambda}}(\Omega_{\rho},\Omega_{\lambda},\xi)=\gamma(\gamma+1)Y_{[\gamma],l_{\rho},l_{\lambda}}(\Omega_{\rho},\Omega_{\lambda},\xi).
\end{equation}

Where $\gamma$ is the grand angular quantum number given by $\gamma=2n+l_\rho+l_\lambda$  ; $l_\rho$ and $l_\lambda$  are the angular momenta corresponding to the Jacobi coordinates $\rho$ and $\lambda$  and $n$ is a non-negative integer number. 

 In the hypercentral approach, the three-quark interaction is assumed to be hypercentral, that is the energy of the system depends only on the quarks distance
\begin{equation}   \label{V3q}
  V_{3q}(\rho,\lambda)=V(x),
\end{equation}
as a consequence, the three-quark wave function is factorized 
\begin{equation}   \label{Psi9x)}
\psi_{3q}(\rho,\lambda)= \psi_{\nu,\gamma}(x)Y_{[\gamma],l_{\rho},l_{\lambda}}(\Omega_{\rho},\Omega_{\lambda},\xi).  
\end{equation}
where $\nu$ determines the number of the nodes of the wave function. The hyperradial wave funtion $ \psi_{\nu,\gamma}(x) $ is obtained as a solution of the hyperradial equation

\begin{equation}\label{Schro 2}
[\frac{d^2}{dx^2}+\frac{5}{x}\frac{d}{dx}-\frac{\gamma(\gamma+4)}{x^2}]\psi_{\nu,\gamma}(x)=-2m[E_{\nu,\gamma}
-V(x)]\psi_{\nu,\gamma}(x).
\end{equation}

\subsection{Potential model}

To make a phenomenological model, we should introduce a potential model such that the QCD concepts of the quark-quark interactions be satisfied. From the experimental observations, we find that all hadrons are made of quarks and no single quark can be individually observed. This fact imply that the quarks are confined in hadrons in the vacuum. According to quantum chromodynamics, there are super-strong color attractive interactions among the quarks, causing three quarks of different colors to be confined together and form a colorless baryon. Moreover the experimental baryon spectroscopy shows an underlying SU(6) symmetry. Therefore an interaction potential should contain two main terms: a confining SU(6) invariant term, and a SU(6) breaking term describing the splitting of multiplets of baryons. Thus the three-quark interaction can be generally written in the form of

 \begin{equation}
V_{3q}=V_{SU(6)-invariant}+V_{SU(6)-breaking}+V_{Spin-Orbit}.
\end{equation}

for the $V_{SU(6)-invariant}$ sector, the Coulomb-plus-linear potential ($-\frac{\tau}{x}+\beta x$), 
known as the Cornell potential, has received a great deal of attention both in particle and atomic physics. Our potential model is a combination of lattice QCD calculations plus the Isgur-Karl interaction \cite{Karl 1, Isgur 1}.  Our potential includes the short distance Coulombic interaction of quarks and the large distance quark confinement via the linear terms in a simple form. Coulombic term alone is not sufficient because it would allow free quarks to ionize from the system. The nonperturbative SU(6)-invariant part is introduced as follows:

 \begin{equation}  \label{confining}
V_{SU(6)-invariant}=-\frac{\tau}{x}+\beta x
\end{equation}

where $x$ is the hyper radius  and $\tau$and $\beta$ are constant. The perturbative interactions are considered as a combination of SU(6)-breaking and spin-orbit interaction terms.  The spin-orbit interaction is introduced as 

\begin{equation} \label{spin-orbit}
V_{Spin-Orbit}=V_{\gamma s}(\overrightarrow{\gamma}. \overrightarrow{S})
\end{equation}

where

\begin{equation}
V_{\gamma s}(x)=\frac{1}{2m_i m_j x}[3\frac{dV_V}{dx}-\frac{dV_S}{dx}].
\end{equation}

$ V_V $ and $ V_S $ are the coulombic and confining parts of Eq. \ref{confining} respectively, and $ m_i $ and $ m_j $ are the masses of  $i$th and $j$th quarks. The SU(6)-breaking part of the interactions is considered as follows:

\begin{equation}
V_{SU(6)-breaking}=V_{Spin}+V_{Isospin}.
\end{equation}

 We make the following model for spin-spin and isospin-isospin interaction

\begin{equation}   \label{Hs}
V_{Spin}=\frac{A_{S}}{6m_\rho m_\lambda}\frac{e^{\frac{-x}{x_{0S}}}}{x x_{0S}^2}\Sigma_{i<j}(\overrightarrow{s_{i}}.\overrightarrow{s_j}),
\end{equation}
and
\begin{equation}  \label{Hi}
V_{Isopin}=\frac{A_{I}}{6m_\rho m_\lambda}\frac{e^{\frac{-x}{x_{0I}}}}{x x_{0I}^2}\Sigma_{i<j}(\overrightarrow{t_{i}}.\overrightarrow{t_j}), 
\end{equation}

where $\overrightarrow{s_i}$ and $\overrightarrow{t_i}$ the spin and isospin operators of the $i$ th
quark respectively and $ A $ and $ x_0 $ are constants. Then from Eqs. (\ref{spin-orbit}-\ref{Hi}), the hyperfine interaction is given by
\begin{equation}
V_{hyp}=V_{Spin}+V_{Isospin}+V_{spin-Orbit}.
\end{equation}
The contributions of this hyperfine interaction is added to the unperturbed SU(6)-invariant energies provided by the potential \ref{confining}. In the next section, we obtain the wave function and energy of the system in the 
framework of a simple approximation with confining potential \ref{confining}.\\

\subsection{Solution of Schr\"{o}dinger equation}

The equation \ref{Schro 2} can be solved analytically for the hyperCoulomb potential 
\begin{equation}\label{coulomb}
V_{hc}=-\frac{\tau}{x}.
\end{equation}

The eigenvalues of the hyperCoulomb problem can be obtained by generalizing to six dimensions. The calculations performed in three dimensions, obtaining 

\begin{equation}\label{Ecoulomb}
E_{n,\gamma}=-\frac{\tau^2m}{2n^2}
\end{equation}
where $ n=\gamma+\frac{5}{2}+\nu $ is the principal quantum number and $ \nu=0,1,2, ... $ is the radial quantum number that counts the number of nodes of the wave function.\\
The eigenfunctions of \ref{Schro 2} with the hyperCoulomb potential can be obtained analytically and are \cite{Giannini:307}

\begin{equation}\label{psi}
\psi_{\nu,\gamma}(x)=[\frac{\nu!(2g)^6}{(2\gamma+2\nu+5)(2\gamma+\nu+4)!}]^{\frac{1}{2}} (2gx)^\gamma e^{-gx}L^{2\gamma+4}_{\nu}(2gx)
\end{equation}

where $ g=\frac{m\tau}{\gamma+\nu+5/2} $ and $ L^{2\gamma+4}_{\nu}(2gx) $ are the Laguerre polynomials. 
We have plotted the distribution of the hyperradial wave function density for the ground and first radial excited states  of the $ qqc $ state baryons in fig. 1.\\

The hyperCoulomb potential does not confine quarks in hadrons and therefore we add the linear confining potential to the hyperCoulomb interaction \ref{coulomb} and therefore the potential we use is in the form of the equation \ref{confining}. The equation  can not be solved analytically with potential \ref{confining}, except when the linear term is treated as a perturbation. Here, we consider this situation and using the obtained wave function \ref{psi} we can obtaine the expectation values of the linear term for different baryon states. Therefore the energy eigenvalues for the baryons are obtained as

\begin{equation}\label{Etotal}
E_{n,\gamma}=-\frac{\tau^2m}{2n^2}+\frac{\beta}{2m\tau}[3n^2-\gamma(\gamma+4)-\frac{15}{4}]
\end{equation}

which is valid, to a good approximation, for the low lying states.

\section{The spectrum and electromagnetic transitions}

\noindent Using the obtained energy eigenvalues \ref{Etotal} and wave function \ref{psi} one can calculate the baryon mass by sum of the quark masses, energy eigenvalues and the hyperfine interaction potential treated as a perturbation:

\begin{equation}\label{Mass}
M_{baryon}=\sum_{i=1}^{3}m_i-\frac{\tau^2m}{2n^2}+\frac{\beta}{2m\tau}[3n^2-\gamma(\gamma+4)-\frac{15}{4}]+<H_{hyp}>
\end{equation}

where using the unperturbed wave function \ref{psi} we can obtain the perturbative hyperfine interaction $  <H_{hyp}> $. The parameters used in our model (listed in table \ref{tab:parameters}) are obtained from our previous work \cite{Ghalenovi:26} in which we have studied the light baryon resonances in a quark model. We use the experimental mass of $ \Xi_{cc}(3520) $ baryon to determine the mass of the charm quark. The value of $ A_{I} $ is also fitted to the mass difference between $ \Sigma_c(2455) $  and $ \Lambda_c(2286) $ states. \\
Our predictions for single charm baryon spectra are listed in tables \ref{tab:mass1}-\ref{tab:mass2} and compared with the experimental data \cite{PDG:2018, CLEO:2001} or other theoretical predictions \cite{Shah:123102,  Ebert:014025, Yoshida:114029,Chen:82, Yamaguchi:2034034817}. For $ \Lambda_c^+ $ state, as the first discovered singly charmed baryon, our results are in good agreement with those of other works. The experimnetal known ground state is $ \Lambda_c(2286)^+ $ and the first orbital excited states are  $ \Lambda_c(2595)^+ $  and  $ \Lambda_c(2625)^+ $ with quantum numbers $J=1/2^{-}$ and $J=3/2^{-}$  respectively. $ \Lambda_c(2765)^+ $ was also observed by CLEO collaboration \cite{CLEO:2001} but its quantum number is still unknown.  Our predictions indicate that $ \Lambda_c(2765)^+ $ can be considered as the first radial excited state of $ \Lambda_c(2286)^+ $ with $ J^P=1/2^{+} $ which is in good agreement with other predictions \cite{Shah:123102,  Ebert:014025, Chen:82, Yamaguchi:2034034817}. $ \Lambda_c(2880)^+ $   with $J=5/2^{+}$ is also known and our results show it can be considered as the second orbital excited state of $ \Lambda_c(2286)^+ $ in favor of Refs. \cite{Shah:123102,  Ebert:014025,Chen:82}.\\                    
For $ \Sigma_c $, two ground states $ \Sigma_c(2455) $ and $ \Sigma_c(2520) $ with $ J^P $ values $ J=\frac{1}{2}^+ $ and $ J=\frac{3}{2}^+ $ has reported by Ref. \cite{PDG:2018}. $ \Sigma_c(2800) $ is also found but its quantum number is still unknown. Refs. \cite{Shah:123102,Ebert:014025} have predicted that as $ 1P $  state.  Our results listed in table \ref{tab:mass2} indicate that we can consider $ \Sigma_c(2800) $  as  $ 1P $ state with a mass difference of 62 $ MeV $  to the experimental mass  $ 2806 MeV $. Our results show that $ 2S $ state is also a good candidate for $ \Sigma_c(2800) $.\\

Within the baryons the mass of the quarks may get modified due to its binding interactions with the other two quarks. The effective quark mass is defined as 

\begin{equation}\label{parameters}
m_{i}^{eff}=m_{i}(1+\frac{E_{n,\gamma}+<H_{hyp}>}{\sum_{i}m_{i}})
\end{equation}

Such that the mass of the baryon is

\begin{equation}\label{Mass2}
M_{baryon}=\sum_{i}m_{i}^{eff}
\end{equation}

The magnetic moment of baryon is obtained in terms of the spin-flavour wave function of the constituent quarks as

\begin{equation}\label{Mu}
\mu_{B}=\sum_{i}\left\langle \phi_{sf}|\mu_{i}\vec{\sigma}_{i}|\phi_{sf}\right\rangle
\end{equation}

where $ \mu_{i}=\frac{e_i}{2m_{i}^{eff}} $. Here, $e_i$ and $s_i=\frac{\sigma_i}{2}$ represent the charge and the spin  of the quark  constituting the baryonic state and $|\phi_{sf}>$ represents the spin-flavour wave function of the respective baryonic state. The magnetic moments of charm baryons obtained in our model are listed in table \ref{tab:Mu1} and compared with other predictions \cite{ Patel:065001, Sharma:073001, Ghalenovi:20, Kumar:141, Faessler:094013}. The present calculations are in very good agreement with those of Ref. \cite{Kumar:141} and our previous work \cite{Ghalenovi:20}.\\
We can obtain the magnetic moments of different transitions  by sandwiching $\sum_{i}\mu_{i}\vec{\sigma}_{i}$ between the appropriate $3/2^{+}$ and $1/2^{+}$ baryon wave functions. The transition magnetic moments for $\frac{3}{2}^{+}\rightarrow\frac{1}{2}^{+}$ are computed as \cite{Ghalenovi:57, Ghalenovi:133}

\begin{equation}\label{Mu transition}
\mu_{\frac{3}{2}^{+}\rightarrow\frac{1}{2}^{+}}=\Sigma_{i} <\phi_{sf}^{\frac{3}{2}^{+}}|\mu_{i}\sigma_i|<\phi_{sf}^{\frac{1}{2}^{+}}>.
\end{equation}

In order to evaluate $B_{3/2}\rightarrow B_{1/2}\gamma$  transition magnetic moments, we take the geometric mean of effective quark masses of the constituent quarks of initial- and final-state baryons:
\begin{equation}
m^{eff}=\sqrt{m^{eff}_{i(\frac{3}{2}^{+})}m^{eff}_{i(\frac{1}{2}^{+})}}
\end{equation}

where $m^{eff}_{i}$ is the effective quark mass of the ith quark inside the corresponding baryon.\\

Our results for the transition magnetic moments  are listed in table \ref{Tr MM} and compared with other results  \cite{Sharma:073001, Dhir:243, Bernotas:074016, Aliev:056005}. Our predictions are very close to those obtained by Ref. \cite{Dhir:243}. \\

\section{Conclusions}
In this paper we have investigated the non-strange charm baryons by employing the hyperspherical formalism with a color Coulomb plus linear confining potential. Using the theory of time-independent perturbation for the hyperfine interactions, we got the effects of spin, isospin and spin-orbit potentials in the shift of baryon energy. We have considered the color Coulomb part as  parent and the linear term as a perturbation. This method is valid, to a good approximation, for the low lying state baryons. Using the hyperCoulomb wave function we could obtain the mass spectra and transition magnetic moments of single charm baryons. Our results are in good agreement with those of other works.

\section{Acknowledgments}
 This work was supported by Kosar University of Bojnord with the Grant number (No. 9809141444).

\newpage

 \begin{table} [ptb] 
 \small
 
 \caption{Quark-model parameters}
 \label{tab:parameters}
 \begin{center}
\begin{tabular}{l|l} \hline  \hline            
$m_{q}$  & 313 $MeV$\\ 
$m_{c}$  & 1747 $MeV$\\
$\beta$  & 0.87 $ fm^{-1}$\\
$\tau$ & 2.54 \\
$A_S$&67.4 \\
$x_{0S}$& 1.93 $fm$\\
$A_I$ & 390.2 \\
$x_{0I}$ & 1.87 $fm$\\
\end{tabular}
  \end{center}
   \end{table}

\begin{figure}[hbtp]
\centering
\includegraphics[scale=0.6]{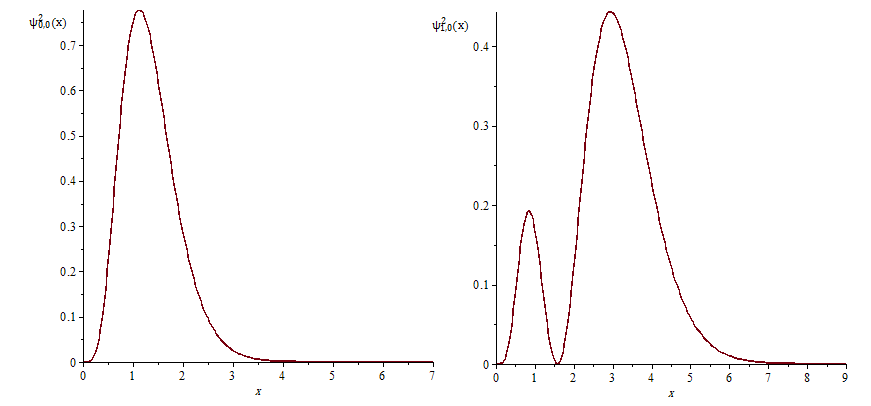}
\caption{$ \psi_{\nu,\gamma}^2 $ for the ground state (left) and first radial excited state (right) vs. $ x $.}
\end{figure}

\begin{table}   
\small
\caption{Mass spectra of $\Lambda_{c}^+ $ baryon (in MeV).}
\label{tab:mass1}
 \begin{center}
{\begin{tabular}{lllllll} \hline \hline   
$ n^{2S+1}L_J $  & Our&Exp& Ref.\cite{Shah:123102} & Ref.\cite{Ebert:014025}  &Ref.\cite{Chen:82} &  Ref.\cite{Yamaguchi:2034034817}\\  \hline

$1^2S_{\frac{1}{2}}$   &2285  &2286.46$\pm$0.14 &2287 &2286 &2286 &2268\\
$2^2S_{\frac{1}{2}}$   & 2740 &2766$\pm$2.4 &2758 &2769 &2766 &2791\\
$3^2S_{\frac{1}{2}}$   &3176  & &3134 &3130 &3112 &2983\\ 
$4^2S_{\frac{1}{2}}$   &3672  & & 3477&3430 &3397 &3154\\\hline

$1^2P_{\frac{1}{2}}$   &2693  &2592.25$\pm$0.28 &2694 &2589 &2591 &2625 \\
$1^2P_{\frac{3}{2}}$   &2651 &2628.11$\pm$0.19 &2640 &2627 &2629 &2830\\ \hline

$2^2P_{\frac{1}{2}}$   &3115& &3062 &2983 &2989 &\\
$2^2P_{\frac{3}{2}}$   &3093& &3015 & 3005&3000 &\\ \hline

$3^2P_{\frac{1}{2}}$   &3598& & 3397&3303 &3296 &\\
$3^2P_{\frac{3}{2}}$   &3591& &3354 &3222 &3301 &\\ \hline

$1^2D_{\frac{3}{2}}$   &2997& &2924 & 2874 &2857 &3120\\
$1^2D_{\frac{5}{2}}$   &2986&2881.63$\pm$0.24 &2854 & 2880&2879 &3125\\ \hline

$2^2D_{\frac{3}{2}}$   &3495& & 3263&3189 &3188 &3194\\
$2^2D_{\frac{5}{2}}$   &3483& &3204 &3209 &3198 &3194\\ \hline

$1^2F_{\frac{5}{2}}$   &3358& & 3130&3097 &3075 &3092\\
$1^2F_{\frac{7}{2}}$   &3344& &3052 &3078 & 3092&3128\\ \hline
\end{tabular}}
 \end{center}
\end{table}

\begin{table}  
\small 
\caption{Mass spectra of $ \Sigma_{c}^0 $ baryon (in MeV).}
\label{tab:mass2}
 \begin{center}
{\begin{tabular}{lllllll} \hline \hline  
$ n^{2S+1}L_J $ &  Our&Exp&  Ref.\cite{Shah:123102}  & Ref.\cite{Ebert:014025} &Ref.\cite{Yoshida:114029} &  Ref.\cite{Yamaguchi:2034034817}\\  \hline

$1^2S_{\frac{1}{2}}$   &2452  &2453.75$\pm$0.14 & 2452& 2443& 2460&2455\\
$2^2S_{\frac{1}{2}}$   & 2801 & &2891 &2901 &3029 &2958\\
$3^2S_{\frac{1}{2}}$   & 3202 & & 3261& 3271&3171 &3115\\ 
$4^2S_{\frac{1}{2}}$   & 3685 & & 3593&3581 & &\\\hline

$1^4S_{\frac{3}{2}}$   &2510  &2518.48$\pm$0.2 &2518 &2519 &2523 &2519 \\
$2^4S_{\frac{3}{2}}$   &2815 & &2917 &2936 &3065 &2995\\
$3^4S_{\frac{3}{2}}$   & 3210& & 3274& 3293&3094 &3116\\
$4^4S_{\frac{1}{2}}$   &  3689& &3601 & 3598& &\\\hline

$1^2P_{\frac{1}{2}}$   &2744& 2806$^{+5}_{-7}$& 2809&2799 & 2802&2848\\
$1^2P_{\frac{3}{2}}$   &2702& &2755 &2798 &2807 &2763\\
$1^4P_{\frac{1}{2}}$   &2803& &2835 & 2713& &\\
$1^4P_{\frac{3}{2}}$   &2761& & 2782& 2773& &\\
$1^4P_{\frac{5}{2}}$   &2691& &2710 &2789 &2839 &2790\\ \hline

$2^2P_{\frac{1}{2}}$   & 3136& &3174 &3172 &2826 &\\
$2^2P_{\frac{3}{2}}$   &3115& &3128 &3172 &2837 &\\
$2^4P_{\frac{1}{2}}$   &3164& & 3196&3125 & &\\
$2^4P_{\frac{3}{2}}$   &3143& &3151 &3151 & &\\
$2^4P_{\frac{5}{2}}$   &3108& &3090 & 3161&3316 &\\ \hline

$3^2P_{\frac{1}{2}}$   &3615 & &3505 &3488 &2909 &\\
$3^2P_{\frac{3}{2}}$   &3602& &3465 &3486 &2910 &\\
$3^4P_{\frac{1}{2}}$   &3631& & 3525&3455 & &\\
$3^4P_{\frac{3}{2}}$   &3618& &3485 & 3469& &\\
$3^4P_{\frac{5}{2}}$   &3597& &3433 & 3475&3521 &\\ \hline

$1^2D_{\frac{3}{2}}$   &3023& &3112 & 3043 & &3095\\
$1^2D_{\frac{5}{2}}$   &3003& & 2993 &3038  &3099 &3003\\ 
$1^4D_{\frac{1}{2}}$   &3054& &3036  & 3041 & &\\
$1^4D_{\frac{3}{2}}$   &3041& & 3061 & 3040 &  & \\
$1^4D_{\frac{5}{2}}$   &3020& & 2968 & 3023 & &\\
$1^4D_{\frac{7}{2}}$   &2991& & 2909& 3013& &3015\\ \hline

$2^2D_{\frac{3}{2}}$   &3512& &3398 &3366 & &\\
$2^2D_{\frac{5}{2}}$   &3504& &3316 &3365 & 3114&\\
$2^4D_{\frac{1}{2}}$   &3522 & &3376 &3370 & &\\
$2^4D_{\frac{3}{2}}$   &3515& &3442 & 3364&  &\\
$2^4D_{\frac{5}{2}}$   &3499& &3339 &3349 & &\\
$2^4D_{\frac{7}{2}}$   &3484& &3265 &3342 & &\\ \hline

$1^2F_{\frac{5}{2}}$   &3363& & 3245 & 3283 & &\\
$1^2F_{\frac{7}{2}}$   &3349& & 3165 & 3227 & &\\
$1^4F_{\frac{3}{2}}$   &3380& & 3332 & 3288 & &\\
$1^4F_{\frac{5}{2}}$   &3370& & 3268 & 3254 & &\\
$1^4F_{\frac{7}{2}}$   &3356& & 3189 & 3253 & &\\
$1^4F_{\frac{9}{2}}$   &3338& &3094 &3209 & &\\ \hline

\end{tabular}}
 \end{center}
\end{table}

\begin{table}
\small
\caption{Magnetic moments of the ground states of  $ \Sigma_{c} $ and $ \Lambda_{c}^0 $  baryons in terms of  $ \mu_{N} $ . \label{tab:Mu1}}
 \begin{center}
{\begin{tabular}{llllllll} \hline \hline  
Baryon &$ \mu_B $& Prediction  & Ref.\ \cite{ Patel:065001}&Ref.\ \cite{Sharma:073001}    &Ref.\  \cite{Ghalenovi:20} & Ref.\ \cite{Kumar:141}& Ref.\ \cite{Faessler:094013} \\  \hline
 $\Lambda_{c}^{+}$&$ \mu_c $ &0.37& 0.38&0.392 & &0.37 & 0.37 \\
$\sum_{c}^{++}$  &$ \frac{4}{3}\mu_u-\frac{1}{3}\mu_c $   & 2.45& 2.27&2.20  &&2.36 &1.86 \\
$\sum_{c}^{+}$    &$ \frac{2}{3}\mu_u+\frac{2}{3}\mu_d-\frac{1}{3}\mu_c $  &0.52 &0.50  &0.30 & &0.62 &  0.37 \\
$\sum_{c}^{0}$    &$ \frac{4}{3}\mu_d-\frac{1}{3}\mu_c $  &-1.39 & -1.01&-1.60 &  &-1.36 &-1.11 \\
$\sum_{c}^{*++}$ &$ 2\mu_u+\mu_c $   &4.10&3.84 & 3.92  & 4.10 & &   \\ 
$\sum_{c}^{*+}$   &$ \mu_u+\mu_d+\mu_c $ &1.27&1.25&0.97 &  1.32 & &   \\ 
$\sum_{c}^{*0}$   &$ 2\mu_d+\mu_c $  &-1.54&-0.84 &-1.99 &  -1.44 & &   \\ \hline
\end{tabular}}
\end{center}
\end{table}

\begin{table} 
\small
    \caption{ Transition Magnetic moments $(|\mu_{\frac{3}{2}^{+}\rightarrow\frac{1}{2}^{+}}|)$ of single charm baryons in $\mu_{N}$.}  \label{Tr MM}

    \centering
    \begin{center}
    \begin{tabular}{ccccccc}
    \hline
     Decay Mode&$ \mu_{B\rightarrow B^\prime \gamma}$ &Our&  Ref.\ \cite{Sharma:073001} &Ref.\  \cite{Dhir:243}  & Ref.\ \cite{Bernotas:074016} &Ref.\ \cite{Aliev:056005} \\
\hline
$\Sigma^{*++}_{c}\rightarrow\Sigma^{++}_{c}\gamma$ &$ \frac{2\sqrt{2}}{3} (\mu_u-\mu_c)$ &1.47 &1.37&1.41  &0.90 & 1.33$ \pm0.38 $ \\

$\Sigma^{*+}_{c}\rightarrow\Sigma^{+}_{c} \gamma$ &$ \frac{\sqrt{2}}{3} (\mu_u+\mu_d-2\mu_c)$ &0.12&0.003 &0.09 &0.06 &0.57$ \pm0.09 $ \\

$\Sigma^{*0}_{c}\rightarrow\Sigma^{0}_{c}\gamma$&$ \frac{2\sqrt{2}}{3} (\mu_d-\mu_c)$ &1.21 &2.40 &1.22& 1.03&0.24$ \pm0.05 $ \\

$\Sigma^{*+}_{c}\rightarrow\Lambda^{+}_{c}\gamma$& $ \frac{\sqrt{2}}{\sqrt{3}} (\mu_u-\mu_d)$&2.41& 1.48 &2.28 &1.70 &2.00$ \pm0.53 $\\

\hline
    \end{tabular}
    \end{center}
 
\end{table}


\bigskip

\begin{thebibliography}{31}


\bibitem {Shah:42}
Z. Shah and A. Kumar Rai, Chin. Phys. C \textbf{42}, 5 (2018).

\bibitem {Ghalenovi:2013}
Z.~Ghalenovi et al., Chin. J. Phys. \textbf{51}, 6 (2013).

\bibitem{Ebert5}
D. Ebert, R. N. Faustov and V. O. Galkin, Phys. Rev. D \textbf{84}, 014025 (2011).

\bibitem{Mathur}
N. Mathur, R. Lewis, and R. M. Woloshyn, Phys. Rev. D \textbf{66}, 014502 (2002).

\bibitem{Mao} 
Q. Mao, H. X. Chen, W. Chen, A. Hosaka, et al., Phys. Rev. D \textbf{92}, 114007 (2015).

\bibitem{Liu3} 
Liu et al., Phys. Rev. D \textbf{77}, 014031 (2008).

\bibitem{Aliev6} 
T. M. Aliev, K. Azizi, A. Ozpineci, Nucl. Phys. B \textbf{808}, 137(2009).

\bibitem{Aliev7} 
T. M. Aliev, K. Azizi, and A. Ozpineci, Phys. Rev. D \textbf{79}, 056005 (2009).

\bibitem{Ebert6}
D. Ebert, R. N. Faustov, and V. O. Galkin, Phys. Rev. D \textbf{72}, 034026 (2005).

\bibitem{Thakkar6}
K. Thakkar, A. Majethiya and P. C. Vinodlumar, Eur. Phys. J. Plus \textbf{131}, 339 (2016).

\bibitem{Thakkar7}
A. Majethiya, K. Thakkar and P. C. Vinodlumar, Chin. J. Phys. \textbf{54}, 495 (2016).

\bibitem{Gerasyuta}
 S. M. Gerasyuta, and D. V. Ivanov, Nuovo Cimento A \textbf{112}, 261 (1999).
 


\bibitem {Karl 1}
N. Isgur and Karl, Phys. Rev. D \textbf{19}, 2653 (1978).

\bibitem{Isgur 1}
N. Isgur and G. Karl, Phys. Rev. D \textbf{18}, 4187 (1978).



\bibitem{Giannini:307}
E. Santopinto, F. Iachello and M. M. Gainnini, Eur Phys. J. A \textbf{1}, 307 (1998).

\bibitem{Ghalenovi:26}
Z. Ghalenovi and M. Moazzen Sorkhi, Int. J. Mod. Phys. E \textbf{26}, 7 (2017).

\bibitem{PDG:2018}
M. Tanaboshi et al (Particle Data Group), Phys. Rev. D \textbf{98}, 030001 (2018).


  \bibitem{CLEO:2001}
  M. Artiro et al (CLEO collaboration), Phys. Rev. Lett. \textbf{86}, 4479 (2001).

  \bibitem{Shah:123102} 
Z. Shah, K. Thakkar, A. Kumar Rai and P. C. Vindokumar, Chin. Phys. C \textbf{12}, 123102 (2016)


\bibitem{Ebert:014025}
D. Ebert, R. N. Faustov, and V. O. Galkin, Phys. Rev. D \textbf{84}, 014025 (2011).


  \bibitem{Yoshida:114029}
T. Yoshida, E. Hiyama, A. Hosaka, M. Oka, and K. Sadato, Phys. Rev. D \textbf{92}, 114029 (2015).

    \bibitem{Chen:82} 
B. Chen, K. W. Wei, and A. Zhang, Eur. Phys. J. A \textbf{51}, 82 (2015).


  \bibitem{Yamaguchi:2034034817}

Y. Yamaguchi, S. Ohkoda, A. Hosaka, T. Hyodo, and S. Yasui, Phys. Rev. D \textbf{91}, 034034 (2015).
 
 \bibitem{Patel:065001}
 B.Patel, A. Kumar Rai and P.C. Vinodkumar, J. Phys. G: Nucl. Part. Phys. \textbf{35}, 065001 (2008).
 
  
  \bibitem{Sharma:073001} 
N. Sharma, B. Dahiya, P.K. Chately and M. Gupta, Phys. Rev. D \textbf{81}, 073001 (2010).
 \bibitem{Ghalenovi:20} 
 Z. Ghalenovi, A.A. Rajabi, S. Qin and D.H. Rischke, Mod. Phys. Lett. A \textbf{29}, 20 (2014).
 
 
 \bibitem{Kumar:141}
 S. Kumar, R. Dhir and R.C. Verma, J. Phys. G: Nucl. Part. Phys. \textbf{31}, 141 (2005).

 \bibitem{Faessler:094013}  
 A. Faessler et al., Phys. Rev. D \textbf{73}, 094013 (2006).
 
 
 \bibitem{Ghalenovi:57} 
Z. Ghalenovi, Int. J. Theor. Phys.  \textbf{57}, 9 (2018).

 \bibitem{Ghalenovi:133} 
Z. Ghalenovi and M. Mozzaen Sorkhi, Eur. Phys. J. Plus \textbf{133}, 301 (2018).


  \bibitem{Dhir:243} 
R. Dhir and R.C. Verma, Eur. Phys. J. A \textbf{42}, 243 (2009).


 \bibitem{Bernotas:074016}  
 A. Bernatos and V. Simonis, Phys. Rev. D \textbf{87}, 074016  (2013).
 
 
  \bibitem{Aliev:056005}  
 T. M. Aliev, K. Azizi, and A. Ozpineci, Phys. Rev. D \textbf{79}, 056005 (2009).
 
 
\end{thebibliography}
\end{document}